\documentclass[12pt]{article}
\usepackage{graphicx}

\makeatletter

\renewcommand{\section}{\@startsection {section}{1}{\z@}
{-3.5ex plus -1ex minus -.2ex}{2.3ex plus .2ex}{\normalsize\bf}}

\renewcommand{\subsection}{\@startsection{subsection}{2}{\z@}
{-3.25ex plus -1ex minus -.2ex}{1.5ex plus .2ex}{\normalsize\it}}

\def\abstract{\if@twocolumn
\section*{\abstractname}
\else \small
\quotation
\fi}
\def\endabstract{\if@twocolumn\else\endquotation\fi}

\renewcommand{\@makefnmark}{\hbox{\mathsurround=0pt
$^\dagger$}}
\renewcommand{\@makefntext}[1]{\parindent=1em\noindent
\hbox to 1.8em{\hss$^\dagger$}#1}

\def\thebibliography#1{\section*{\refname\@mkboth
 {\uppercase{\refname}}{\uppercase{\refname}}}\list
 {[\arabic{enumi}]}{\settowidth\labelwidth{[#1]}\leftmargin\labelwidth
 \advance\leftmargin\labelsep\parsep=0em\itemsep=0em
 \usecounter{enumi}}
 \def\newblock{\hskip .11em plus .33em minus .07em}
 \sloppy\clubpenalty4000\widowpenalty4000
 \sfcode`\.=1000\relax}

\makeatother

\begin{document}

\begin{center}

{\normalsize\bf
 DOES GIANT MAGNETORESISTANCE SURVIVE IN PRESENCE OF SUPERCONDUCTING CONTACT?} \\
\bigskip
N.~Ryzhanova$^{1,3}$, C.~Lacroix$^1$,
 A.~Vedyayev$^{2,3}$, D.~Bagrets$^{2,3}$, and B.~Dieny$^2$
\medskip \\
{\small\it
$^1$ Centre National de la Rechercher Scientifique, Laboratoire Louis Neel, 38042  Grenoble, France\\
$^2$CEA/Grenoble, D\'epartement de Recherche Fondamentale sur la 
Mati$\grave e$re Condens\'ee,
SP2M/NM, 38054 Grenoble, France \\
$^3$Departement of Physics, M.~V.~Lomonosov Moscow State University,
119899 Moscow, Russia}
\end{center}

\begin{abstract}
The giant magnetoresistance (GMR) of ferromagnetic bilayers with a superconducting contact
(F1/F2/S) is calculated in ballistic and diffusive regimes. As in spin-valve, it is assumed that the
magnetization in the two ferromagnetic layers F1 and F2 can be changed from parallel to antiparallel.
It is shown that the GMR defined as the change of conductance between the two magnetic 
configurations is an oscillatory function of the thickness of F2 layer and tends to an asymptotic 
positive value at large thickness. This is due to the formation of quantum well states in F2 induced
by Andreev reflection at the F2/S interface and reflection at F1/F2 interface in antiparallel
configuration. In the diffusive regime, if only spin-dependent scattering rates in the magnetic
layers are considered (no difference in Fermi wave-vectors between spin up and down electrons) then
the GMR is supressed due to the mixing of spin up and down electron-hole channels by Andreev 
reflection.
\bigskip


\end{abstract}

The mechanisms of Giant Magnetoresistance (GMR) in magnetic multilayers and sandwiches
are well established.  They are related to the influence of the spin assymetry on the
conductivity. Several microscopic factors may play a role: the electronic band structure
(e.g. exchange splitting)~\cite{ref1} has a direct influence whereas the spin assymetry
in the bulk or interface electron scattering rates  has an indirect influence~\cite{ref2}. 
Experiments on GMR are carried out in two geometries: current in plane of the structure
(CIP)~\cite{ref3} and current perpendicular to plane (CPP)~\cite{ref4}.  It was 
observed that the CPP-GMR amplitude is always larger than the CIP one~\cite{ref4}. It should be mentioned that
in CPP measurements, superconducting leads  are most often used as contacts on the spin-valve 
structure~\cite{ref4}. Recently, in Ref.~\cite{ref5} the CPP-GMR of Co/Cu layer was
numerically calculated using a realistic band structure for Co and Cu and in the presence of one
superconducting contact. Unexpectedly, it was found that in this case the GMR is 
completely supressed due to Andreev reflection on  the ferromagnet/superconductor~(F/S) interface.
In order to resolve this contradiction between theory and experiment, we developed an analytical 
theory of CPP-GMR for a  spin valve sandwich of the type F/F/S, where F's are ferromagnetic
layers, the magnetizations of which can be oriented parallel or antiparallel to each other,
S is a superconducting contact. A simple  two band (spin up and down) free 
electron model is adopted for this calculation.

   For calculating the  conductance of the considered system, we used the generalized
Fisher-Lee formulae in spinor form~\cite{ref6}:
\begin{equation}
\sigma^{\alpha\beta\gamma\delta}(z,z') = 
\frac{4e^2}{\pi\hbar}\left(\frac{\hbar^2}{2m}\right)^2\sum_{\kappa} 
A_{\bf\kappa}^{\alpha\delta}(z,z') \stackrel{\leftrightarrow}{\nabla}_z
\stackrel{\leftrightarrow}{\nabla}_{z'}A_{\kappa}^{\beta\gamma}(z',z),  
\end{equation}
where  $\stackrel{\leftrightarrow}{\nabla}_z= \frac12(\stackrel{\rightarrow}{\nabla}_z
-\stackrel{\leftarrow}{\nabla}_z)$ is the antisymmetric gradient operator and \\
$A^{\alpha\beta}_{\kappa}(z,z')=(i/2)(G^{\alpha\beta}_{\kappa}(z,z')-{G^{\beta\alpha}_{\kappa}}^{*}(z',z))$.
Here $G^{\alpha\alpha}$ is the conventional Green function and $G^{\alpha\beta}$
$(\alpha\ne\beta)$ is the anomalous Green function, antisymmetric in spin
indices $({\alpha,\beta})$. The summation is performed over available channels. 
$\kappa=(\kappa_x,\kappa_y)$ is the component of electron momentum in the XY-plane of the layers
and $z$ is the coordinate perpendicular to the XY-plane.

   The Green functions satisfy the system of Gor'kov equations~\cite{ref7}:
\[
(i\omega - (-\frac{\hbar^2}{2m}\frac{\partial^2}{\partial z^2} - \varepsilon_F) + \varepsilon_{ex}(z))G^{\uparrow \uparrow}(z,z')
+\Delta(z)G^{\downarrow \uparrow}(z,z')=\delta(z-z') 
\]
\begin{equation}
(i\omega + (-\frac{\hbar^2}{2m}\frac{\partial^2}{\partial z^2} - \varepsilon_F) + \varepsilon_{ex}(z))G^{\downarrow \uparrow}(z,z')
+\Delta^{*}(z)G^{\uparrow \uparrow}(z,z')=0,   
\end{equation}
where the superconductor gap $\Delta(z)$ is considered as constant inside the superconductor
and zero in the ferromagnetic layers. On the contrary, the exchange splitting parameter 
$\varepsilon_{ex}(z)=\frac{\textstyle \hbar^2}{\textstyle 2m}( {k^{\uparrow}}^2_F - {k^{\downarrow}}^2_F )$
is zero in superconductor and constant in the ferromagnetic layer 
($k^{\uparrow(\downarrow)}_F$ represent Fermi momenta for spin up (down) electrons ). The system of
equations (2) can be solved exactly in the clean limit: e.g. if the thicknesses of the ferromagnetic
layers are much smaller than both the elastic mean free path and the magnetic length 
$\sqrt{\frac{\textstyle \hbar D}{\textstyle \varepsilon_{ex}}}$, where $D$ is the diffusion constant. 
The same assumption was made in the first part
of reference~\cite{ref5}.  

  The expressions of the conductances in parallel~(P) and antiparallel~(AP) alignments of
the magnetization in the adjacent ferromagnetic layers can be written in the following form
\begin{equation}
G^P = \frac{4e^2}{\pi\hbar}\sum_{\kappa} (1-R^2), 
\end{equation}
\begin{equation}
G^{AP} = \frac{4e^2}{\pi\hbar}\sum_{\kappa}\frac{ (1-R^2)(1-r^2)}
{\scriptstyle 1+2r^2 R^2 + r^4 - 2Rr(\cos{2c^{\uparrow}a} - \cos{2c^{\downarrow}a})(1+r^2) - 
2r^2(R^2\cos{2(c^{\uparrow}+c^{\downarrow})a}- \cos{2(c^{\uparrow}-c^{\downarrow})a})},
\end{equation}
where $r=\frac{\displaystyle c^{\uparrow}-c^{\downarrow}}{\displaystyle c^{\uparrow}+c^{\downarrow}}$ 
represents the effective spin
polarization, $R=\frac{\displaystyle c^{\uparrow}c^{\downarrow}-c_2^2}
{\displaystyle c^{\uparrow}c^{\downarrow}+c_2^2}$,
$c^{\uparrow(\downarrow)}= Re\sqrt{{(k_F^{\uparrow\downarrow})}^2-\kappa^2}$, and
$c_2= Re\sqrt{{(k_F^s)}^2-\kappa^2}$. $a$ is the thickness of the intermediate ferromagnetic layer. 
$\kappa_F^s$ is the Fermi wave vector in the superconducting layer. The
upper limit of the sum over $\kappa$ is equal to the minimum value of 
$k_F^{\uparrow(\downarrow)}$ or $k_F^s$. 

   The physical meaning of the obtained expressions 
is rather clear. In the P-configuration, the conductance of the system  decreases compared to
its value  in the  absence of superconducting contact, due to Andreev reflection. This 
conclusion coincides with the result of the numerical calculations of~\cite{ref5}. If $r=0$,( e.g.
the paramagnetic metal in contact with superconductor), expression~(3) coincides with known
results (See eq. (127) in Ref.~\cite{ref8}).
Expression~(4) contains two factors $(1-R^2)$ and $(1-r^2)$. The first one is
due to Andreev reflection on F/S interface and the second one is the usual reflection of electrons
at $F^{\uparrow}$/$F^{\downarrow}$ interface. So, if only these two factors are taken into account,
considering the denominator in (4) equal to unity, a finite GMR amplitude  is still obtained. 
Let's consider then the effect of denominator in expression (4).
It describes the multiple reflections of an electron which  moves inside the ferromagnetic layer
adjacent to the superconductor,  as in a  Fabri-Perro interferometer. These multiple
reflections are responsible for the formation of quantum well states within the layer. As a result,
the conductance $G^{AP}$ is an oscillatory function of the arguments 
$k_F^{\uparrow(\downarrow)}a$, $(k_F^{\uparrow}\pm k_F^{\downarrow})a$, but it never diverges
nor becomes negative. A similar behaviour of conductance was predicted in~\cite{ref9}, for a 
structure composed of a ferromagnetic layer sandwiched between  a superconducting contact on one 
side and a thin oxide barrier on the  other side.

\begin{figure}
\begin{center}
\includegraphics[scale=0.5, angle=-90]{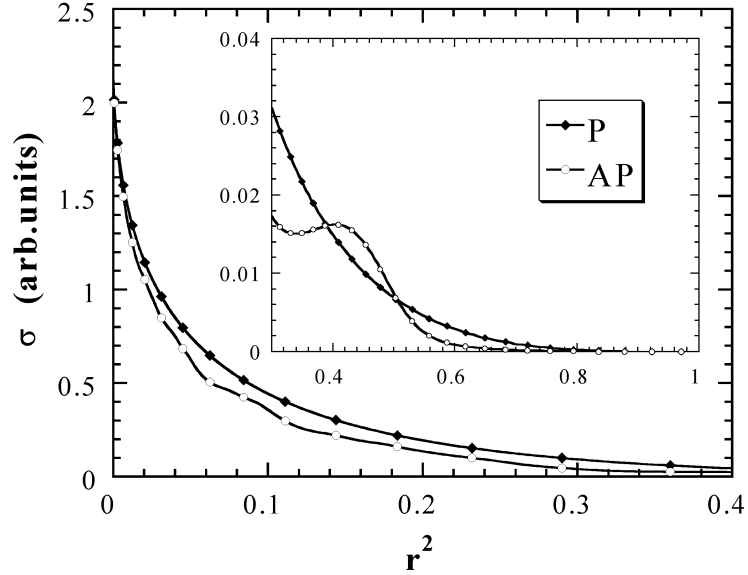}
\end{center}
\caption{The dependencies of conductivities for the  parallel and antiparallel configurations
as a function of the square of the effective spin polarization~$r^2$. The parameters are: $k_F^s=1.3~\AA^{-1}$, 
$a=20.3~\AA$.}
\end{figure}

  Now we come to the question whether $G^P$ is always larger than $G^{AP}$,
or if for some values of parameters, $G^{AP}$ can be equal or even larger then $G^P$. For very small
polarization $r\ll 1$, from expressions (3) and (4), it is easy to obtain the following approximate expression
for the GMR:
\begin{equation}
GMR=\frac{\sigma^P-\sigma^{AP}}{\sigma^{AP}}=2\sum_{\kappa}r^2 R^2 
 =2\sum_{\kappa}\left(\frac{c^{\uparrow}-c^{\downarrow}}{c^{\uparrow}+c^{\downarrow}}\right)^2
\left(\frac{c^{\uparrow}c^{\downarrow}-c_2^2}{c^{\uparrow}c^{\downarrow}+c_2^2}\right)^2  
\end{equation}
This expression is definitely positive.
Without any superconducting contact, the GMR would be given by $GMR=2\sum\limits_{\displaystyle \kappa}r^2$.
It is interesting to note that expression~(5)
coincides with the expression of the MR in a spin-valve  tunnel junction~\cite{ref10}, after 
substitution of $c_2$ by the modulus of the imaginary electron momentum inside the barrier. The physics of
both phenomena is similar: in both cases  the electrons undergo reflections on the 
interface: F/I (I-insulator) or F/S. These spin-dependent reflections change the spin-dependent
density of states in the ferromagnet near the interface and, correlatively change the polarization of the current.
  
  For larger $r$, the conductances $G^P$ and $G^{AP}$ are plotted in fig.1 versus the square of the
effective polarization $r^2$. The following parameters were used: $k_F^{\uparrow}=1\AA^{-1}$, 
$k_F^s=1.3\AA^{-1}$ and $a=5c_0$, ($c_0= 4.06\AA$ is the lattice parameter of Co
for hcp structure).   $a=5c_0$ corresponds to 10 atomic monolayers. 
$k_F^{\downarrow}$ was varied from 1 to 0, so that correspondingly $r^2$ was
changing from 0 to 1. As can be seen in fig.1, the conductances for both magnetic configurations
decrease as  $r^2$  increases. 
$G^{AP}$ exibits also some weak oscillations as a function of $r^2$, but it remains smaller than
$G^P$ for almost the  whole range of $r^2$. For
 $r^2>0.9$, the GMR saturates at a value $\approx 220\%$ (see fig.3) but it does not diverge if $r\rightarrow 1$
as it is the case for a spin-valve without superconducting contact 
($GMR=\frac{\displaystyle 2r^2}{\displaystyle 1-r^2}$ in this case).

\begin{figure}
\begin{center}
\includegraphics[scale=0.5, angle=-90]{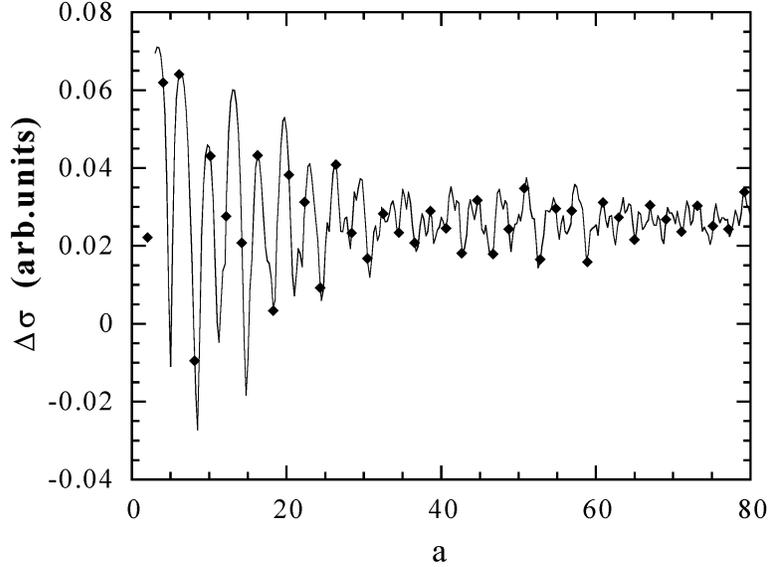}
\end{center}
\caption{The difference of the conductivities $\Delta\sigma = \sigma_P-\sigma_{AP}$
for the  parallel and antiparallel configurations as a function of  the thickness of the ferromagnetic layer $a$.
The parameters are: $k_F^s=1.3~\AA^{-1}$, $k_F^{\uparrow}=1.0~\AA^{-1}$, $k_F^{\downarrow}=0.429~\AA^{-1}$, $r^2=0.16$.
Markers correspond to integer numbers of atomic layers of Co.}  
\end{figure}

 Now let us look at the effect of averaging over a distribution of magnetic layer thickness 
variations since it could arise that these fluctuations may suppress the GMR.  In fig.2,  the difference 
$G^P-G^{AP}$ is plotted versus the thikness $a$ for a given value of $r^2=0.16$. Fig.2 shows that this difference,
and consequently the $GMR=\frac{\displaystyle G^P-G^{AP}}{\displaystyle G^{AP}}$ oscillates around a non-zero
positive value and tends
to the asymptotic limit 44\% for $a>100\AA$. Of course, the thickness of the layer can change only by
steps equal to the lattice parameter $c_0/2$. In fig.2 the possible values of $a$ have been marked 
considering $c_0=4.06\AA$ for Co. It is interesting to note  that the situation is similar to the case of a
spin-valve tunnel junction with paramagnetic metal layer inserted between one ferromagnetic electrode
and an insulating barrier.~\cite{ref11}. In this case, it was shown that the paramagnetic layer (for instance
Cu inserted between Co and Al$_2$O$_3$) can constitute a spin-dependent quantum well. Oscillations in
tunnel magnetoresistance (TMR) were predicted for such system as a function of the paramagnetic layer thickness
with a period given by the Fermi-wave length in this layer. However,
a crucial difference between this  case  and the present one 
is that here, the GMR oscillates around a finite positive value whereas in a tunnel junction, the
GMR oscillates around zero.
Consequently, for  tunnel junctions, averaging over a distribution of the paramagnetic layer thickness caused
by roughness, and/or increasing the paramagnetic layer thickness leads to a strong decay in TMR amplitude.
In contrast, in the present case,  averaging over a distribution
of thickness and/or increasing the thickness of the ferromagnetic layers leads to a non-zero GMR amplitude
which depends on the values $r^2$ and $R^2$. 

This situation is illustrated in fig.3, where the dependence of the GMR on the effective
spin polarization $r^2$ is plotted for two different cases, i.e.
with thickness $a$ equal to 10 and 500 monolayers of Co.
In fig.4, the same dependence is presented but for a structure in which the layer thikness $a$ is supposed to
take random values equally distributed  between 9, 10 or 11 monolayers of Co. 
We have also calculated the $a$-dependence of conductivity  for other values of $r^2$.  These dependences are
similar to the one shown in fig.2. The resonances of $G^{AP}$ exibit rather sharp peaks at 
$a=\frac{\displaystyle 2\pi n}{\displaystyle k_F^{\downarrow}}$ if $1-r^2\ll 1$, i.e
$k_F^{\uparrow}\rightarrow 0$. In this case, the system becomes a  real Fabri-Perro interferometer for electrons. \\

\begin{figure}
\begin{center}
\includegraphics[scale=0.5, angle=-90]{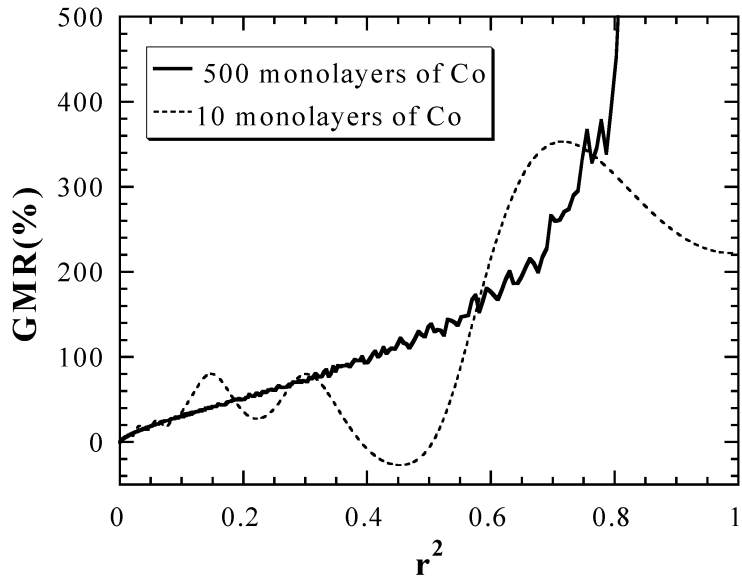}
\end{center}
\caption{The GMR as a function of the square of the effective spin polarization~$r^2$ for
$a=20.3~\AA$~-~dashed line and $a=1015~\AA$~-~solid line.
Parameters are: $k_F^s=1.3~\AA^{-1}$, $k_F^{\uparrow}=1.0~\AA^{-1}$. }  
\end{figure}

\begin{figure}
\begin{center}
\includegraphics[scale=0.5, angle=-90]{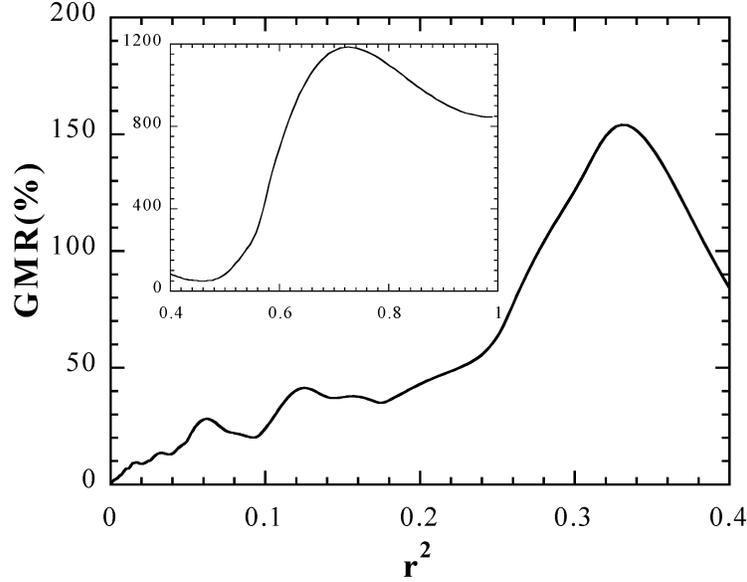}
\end{center}
\caption{The GMR versus the square of the effective spin polarization $r^2$ for
the sandwich with the random (9,10 or 11) numbers of Co monolayers.
Parameters are: $k_F^s=1.3~\AA^{-1}$, $k_F^{\uparrow}=1.0~\AA^{-1}$. }  
\end{figure}

  Now let's consider a different model of the CPP-GMR in spin-valve structures, developed in details
in~\cite{ref12} and often used for the interpretation of CPP-GMR experiments. In this model, it
is considered that charge carriers in ferromagnetic metals are $s$-like electrons with 
negligibly small exchange splitting but with different elastic mean free paths for spin up and 
spin down electrons.  The scattering of $s$-electrons is considered as mainly due to 
$s-d$ scattering, so that the inverse life times of up and down electrons are 
$\frac{\displaystyle \hbar}{\displaystyle \tau_{\uparrow(\downarrow)}}
 = \gamma^2 c\,\nu_d^{\uparrow(\downarrow)}$, where $\gamma$ is the $s-d$
scattering potential, $c$ is the concentration of impurities and 
$\nu_d^{\uparrow}\ne\nu_d^{\downarrow}\gg\nu_s$ are the  densities of states states of $\uparrow$ and
$\downarrow$ spin $d$-electrons, $\nu_s$ is the  $s$-electron density of states. Furthermore, we assume 
that $L\ll l_{sf}$, where $L$ is the thickness of the ferromagnetic layers, $l_{sf}$ is the
spin-flip mean free path of $s$-electrons. Now in Gor'kov equations~(2), $i\omega$ has to be
changed to $i\omega' = i\omega - \gamma^2 c\, G_{dd}^{\uparrow\uparrow(\downarrow\downarrow)}$ and 
$\Delta$ to $\Delta + \gamma^2 c\, G_{dd}^{\uparrow\downarrow}$. It is easy to show that if there 
are no $d$-states in superconductor, then
$G_{dd}^{\uparrow\downarrow} \sim \frac{\displaystyle \gamma^2\nu_d\nu_s}{\displaystyle E_{exch}}
 \ll G_{dd}^{\uparrow\uparrow} \sim {\rho_d}$
in the ferromagnetic layer, due to simultaneous influence of proximity
effect on $G_{ss}^{\uparrow\downarrow}$ and $s-d$ hybridisation. 
Consequently, in the following,  $G_{dd}^{\downarrow\uparrow}$ is neglected.
It is easy to solve Gor'kov equations and calculate the conductance by using expression (1). 
We have to add to (1) vertex correction but, as it was shown in~\cite{ref13}, inclusion of vertex
correction is equivalent to a special choice of effective internal electrochemical fields in such a
manner that the condition $\frac{\displaystyle \partial j}{\displaystyle\partial z}=0$ is satisfied 
($j$ is the current in the $z$-direction). Following this procedure, we found that resistances of the
considered sandwich with superconducting contacts for parallel $R^P$ and 
antiparallel $R^{AP}$ alignment of magnetizations in adjacent F-layers are
equal:
\begin{equation}
R_s^{P}=R_s^{AP}=(a+b)(\rho^{\uparrow}+\rho^{\downarrow})/4 
\end{equation}
where $a$ and $b$ are the thicknesses of the ferromagnetic layers and $\rho^{\uparrow(\downarrow)}$ are
the resistivities for $\uparrow(\downarrow)$ spin $s$-electrons. On the other hand,
if the S-contact is in a normal state, we have
\[
R_N^{P}=\frac{(a+b)\rho^{\uparrow}\rho^{\downarrow}}{\rho^{\uparrow}+\rho^{\downarrow}} 
\]
\begin{equation}
R_N^{AP}=\frac{(a\rho^{\uparrow}+b\rho^{\downarrow})(b\rho^{\uparrow}+a\rho^{\downarrow})}
{(a+b)(\rho^{\uparrow}+\rho^{\downarrow})}  
\end{equation}
Therefore within the assumption that the GMR originates from spin-dependent scattering rates in
the magnetic materials, we find that
there is no GMR effect in presence of superconducting contact. This conclusion
coincides with the results obtained in~\cite{ref14}.

  The absence of GMR in this case can be qualitatively understood  as follows. 
In  a ferromagnetic metal, currents for up and down spin electrons are not
equal. However, in a BCS superconductor, the current is driven by spin-less Cooper pairs, so that up and down
spin currents are equivalent. To maintain this equivalence, electrons undergo  Andreev reflection at the
ferromagnet/superconductor interface and spin accumulation appears at this interface.
Due to this accumulation, a jump $\Delta v$ of chemical potentials (of different signs for up and
down spin electrons) arises. The values of  these jumps are 
$\Delta v^{\uparrow}=v\frac{\displaystyle \rho^{\downarrow} - \rho^{\uparrow}}
{\displaystyle \rho^{\uparrow} + \rho^{\downarrow} } = -\Delta v^{\downarrow}$ for P configuration
and $\Delta v^{\uparrow}=-\Delta v^{\downarrow}=v
\frac{\displaystyle (a-b)(\rho^{\downarrow}-\rho^{\uparrow})}
{\displaystyle (a+b)(\rho^{\uparrow} + \rho^{\downarrow}) }$ for AP configuration, where $v$ 
is the voltage drop across the total structure. In particular,
$\Delta v_{\uparrow}=\Delta v_{\downarrow}=0$ for $a=b$. It can be shown that this additional 
drop of voltage in parallel configuration exactly equalizes the resistances for P and AP configurations,
so that the GMR is suppressed.\\

  In conclusion, contrast to~\cite{ref5}, we have shown that in general, due to the exchange splitting of 
electron bands in ferromagnetic metals,  the presence of a superconducting 
contact adjacent to GMR multilayer does not suppress the GMR amplitude in the CPP geometry,
except for special values of the parameters of the system. 
We think that such a particular  situation has been considered in Ref.~\cite{ref5}. 
The value of GMR depends not only on the spin-polarization of the electrons in the ferromagnetic
layer, but also  on the band parameters in the superconductor (in our case on value of $R^2$). Of course, since
we used a simplified  model, our results have only a qualitive nature. 
A more detailed analyzis of the dependence of the GMR on the  microscopic parameters (exchange splitting, difference in
$\uparrow$ and $\downarrow$ spin mean free paths), taking into account F/S 
interfacial scattering,  will be presented in a forthcoming paper. \\

N.Ryzhanova and A.Vedyayev are grateful to J.Fourier University and  CEA/Grenoble DRFMC/SP2M/NM
for hospitality. The work was partially supported by Russian Foundation for Basic Research, Grant No. 98-02-16806.
\small{

}

\end{document}